\documentclass[a4paper,11pt]{article}
\usepackage{jheppub} 
\usepackage{graphicx,color,rotating}
\usepackage{hyperref}
\usepackage{epsfig,color}
\usepackage{slashed}
\usepackage{amsfonts}
\usepackage{ulem}
\usepackage{color}
\newcommand{\lsim}{\raisebox{-0.13cm}{~\shortstack{$<$ \\[-0.07cm] $\sim$}}~}

\graphicspath{{D:/Desktop/draft}}
\usepackage{tabularx}
\usepackage{graphicx}
\usepackage{adjustbox}
\usepackage{tabularx}
\def\lsim{\:\raisebox{-0.5ex}{$\stackrel{\textstyle<}{\sim}$}\:}

\usepackage{float}
\usepackage{subfig}
\usepackage{amsmath,mathtools}
\begin{document}
	\author{
	Po-Yan Tseng$^{1,2}$ and Yu-Min Yeh$^{1}$}
	\affiliation{
		$^1$ Department of Physics, National Tsing Hua University,
		101 Kuang-Fu Rd., Hsinchu 300044, Taiwan \\
        $^2$ Physics Division, National Center for Theoretical Sciences,
		Taipei 106319, Taiwan \\
	}
	
	\date{\today}

\abstract{The memory-burden effect stabilizes the evaporating Primordial Black Holes (PBHs) before its complete decay. This also suppresses the evaporation flux via the entropy factor to the $k$-th power and circumvents severely astrophysical and cosmological constraints, such that it opens a new mass window for PBH Dark Matter lighter than $10^{15}$ g which has entered the memory-burden phase in the present epoch. In this study, we propose two scenarios to probe PBHs in the earlier semiclassical phase that evaporate at unsuppressed rates. 
The first scenario considers gravitons, emitted semiclassically from PBHs, propagating across the recombination epoch, then the magnetic field in the cosmological filaments converts them into photons via the Gertsenshtein effect.
The second scenario relies on the PBHs mergers today, reproducing young semiclassical black holes with unsuppressed evaporation, but it is highly model dependent and has no sufficient theory support.
For phenomenology studies,
we perform computations of the extragalactic photon spectrum from PBHs emission according to these scenarios. The upper limits on the fractional abundance of PBH are obtained by comparing with the sensitivities of gamma-ray observations. 
The graviton-photon conversion scenario excludes the mass window $7.5\times 10^5\,{\rm g} \leq M_{\rm PBH}\leq 4.4\times 10^7\,{\rm g}$ with $f_{\rm PBH}|_{T_\phi}\geq 1$ and $k=1$, assuming the optimistic magnetic field $B_0=100$ nG.
Meanwhile, the merging scenario, which is insensitive on $k$, restricts PBH Dark Matter lighter than $2.2\times 10^{11}$ g. 
}

\title{Constraining memory-burdened primordial black holes with graviton-photon conversion and binary mergers}

\maketitle

\section{Introduction}
\label{sec:intro}

The hypothesis Primordial black holes (PBHs) are proposed to form in the early Universe and constitute the entire dark matter (DM) abundance~\cite{Hawking:1971ei,Chapline:1975ojl,Khlopov:2008qy,Carr:2016drx,Carr:2020gox,Carr:2020xqk,Green:2020jor}.
This scenario has been thoroughly studied for PBH masses $M_{\rm PBH}$ from Planck mass ($M_{\rm Pl}$) to $10^{20} M_\odot$~\cite{Carr:2020gox}. It leads to a mass window in the asteroid range $M_{\rm PBH}\subset$ [$10^{17}$ g, $10^{22}$ g] where single mass PBH constitute the entirety of dark matter. For PBH lighter than $10^{17}$ g, it is severely constrained due to the black hole (BH) Hawking evaporation~\cite{Hawking:1975vcx,Gibbons:1977mu}. Hawking adopted semi-classical calculation and showed that a BH emits a thermal spectrum of particles, associated with a temperature scale as $T_{\rm PBH} \propto 1/M_{\rm PBH}$. 
Consequently, the PBH evaporation process is self-similar and end with a final burst when $M_{\rm PBH} \to M_{\rm Pl}$.

However, Ref.~\cite{Dvali:2018xpy,Dvali:2020wft} pointed out that the semi-classical approximation will break down when the BH lost about half of its initial mass. Subsequently, the backreaction of the emission on the quantum state of the BH, called the memory burden effect, significantly suppresses the BH evaporation by the $k$-th power of BH entropy $S$.
Therefore, the life history of a PBH can be divided into semiclassical phase (before losing half of its initial mass) and burden phase.
As a result, it prolongs the lifetime of low-mass PBHs and modifies their upper limits of abundance, thus opening up the light-mass window for PBH to be viable dark matter candidate. 
In recent literature, many phenomenological studies have been investigated on burdened PBHs~\cite{Thoss:2024hsr,Chianese:2024rsn,Chaudhuri:2025asm,Tan:2025vxp,Chianese:2025wrk,Loc:2024qbz}.
In particular, for $k=2$, the PBH with initial mass $M_{\rm PBH}\subset$ [$10^{5}$ g, $10^{10}$ g] can explain the entire dark matter abundance
and satisfies various constraints~\cite{Thoss:2024hsr}. This unconstrained mass window is mainly because the detectable photon or neutrino fluxes must be produced after cosmic recombination or matter-radiation equality, which are suppressed by entropy when the PBH is under the burden phase.

Taking advantage of the feeble interaction between the graviton and other Standard Model (SM) particles, the graviton flux produced without entropy suppression from semiclassical phase of PBH, before cosmic recombination, can almost free streaming to present detections.
We might detect those gravitons as a gravitational wave (GW) signal.
Unfortunately, for low-mass PBH interested in memory burden, direct detection of such high-frequency GW signal is beyond the reach of current and future GW interferometers~\cite{Inomata:2020lmk,Ireland:2023avg,Barman:2024ufm}, 
since the energy of gravitons follows the temperature associated with the PBH mass.
However, the Gertseshtein effect converts the gravitons into the photons under the external magnetic field, which significantly improves the detectability of the signal. 
In particular, we used the cosmological filaments as sources of magnetic fields and computed the conversion probability.
This effect 
has been applied to the scenarios that gravitons generated from dark matter decay~\cite{Dunsky:2025pvd,Ramazanov:2023nxz} or PBH evaporation~\cite{Ito:2023nkq}. In this work, we will adopt the latter approach and additionally include the memory burden effect. 
We consider the photon signals from two cases: 
i) The PBH emitted gravitons from both semiclassical and burden phases separated by the PBH formation mass times parameter $q$.
ii) A new PBH from merging of two burdened PBHs restarts the semiclassical phase.
The latter case is, unlike the first, 
insensitive to the parameter $k$, instead depends on $q$. 
Using the sensitivities of current and future gamma-ray observations within energy $E_\gamma\subset$ [$10^{-5}$ GeV, $10^{7}$ GeV], we set the upper bounds for the abundance of PBHs.
This restricts the memory burden scenario so that the monochromatic mass PBH 
lighter than $2.2\times 10^{11}$ g as a viable dark matter candidate.

This paper is organized as follows. In Section~\ref{sec:memory}, we review the memory burden effect and describe the implementation in the BlackHawk v2.3 package. In Section~\ref{sec:graviton_photon}, we formulate the Gertsenshtein effect and take an approximation for the conditions of cosmological filaments.
Then we convolute the graviton spectra from PBH with the conversion probability to obtain the photon flux. 
In Section \ref{sec:pbh_merger}, we briefly introduce the PBH merger and deduce Hawking spectrum from it.
Considering the observational upper limits or sensitivities of extragalactic gamma-ray, we set constraints on PBH abundance in Section~\ref{sec:constraint}. The results are summarized in Section~\ref{sec:summary}.

\bigskip

\section{Memory Burden}
\label{sec:memory}

The memory burden effect considers the quantum state of the black hole (BH), deviating from the prediction of Hawking evaporation which assumes the semiclassical approaches~\cite{Dvali:2018xpy,Dvali:2020wft,Dvali:2024hsb}.
This effect is caused by the information store in the system resists its decay, which
results in slowing down the evaporation and extending the lifetime of the black hole, when a black hole lost a certain fraction of its initial mass and the back reaction becomes significant. 

We follow the calculation of Ref.~\cite{Chianese:2024rsn}.
We denote $M_{\rm PBH}\vert_{T_\phi}$ the initial mass of PBH formed at cosmological temperature $T_\phi$. 
Taking an example of PBHs generated from a dark first-order phase transition (FOPT), $T_\phi$ indicates when the Yukawa interaction dominates a Fermi-Ball (FB) total energy, so that the FB becomes unstable and starts collapsing to a PBH~\cite{Marfatia:2021hcp}.
The Hawking evaporation under the memory burden effect can be divided into two parts: When the mass of PBH is greater than $qM_{\rm PBH}\vert_{T_\phi}$, the PBH is in the semiclassical phase. As evaporation goes on, the mass drops below $qM_{\rm PBH}\vert_{T_\phi}$, the PBH converts into burden phase. In this work, we take $q=0.5$ as the memory burden effect becomes relevant, when BH has lost half of its mass. In the burden phase, the loss rate of PBH mass is suppressed by the entropy factor
    \begin{equation}\label{mass_lossrate}
    \frac{dM_{\rm PBH}}{dt}(t>t_q)=\frac{1}{S(M_{\rm PBH})^k}\frac{dM_{\rm PBH}}{dt},
    \end{equation}
where $k$ is an adjustable parameter that governs the degree of entropy suppression ~\cite{Chaudhuri:2025asm},
and we consider it as a free parameter with $k \ge 0$ in this work ~\cite{Barman:2024iht,Barman:2024ufm,Chianese:2024rsn,Athron:2024fcj}.
PBH entropy and the duration of the semi-classical phase are~\cite{Haque:2024eyh}
    \begin{equation}
    S(M_{\rm PBH})=4\pi G_N M^2_{\rm PBH},\quad t_q=t_{\rm PBH}(1-q^3).
    \end{equation}
For $t > t_q$, the quantum effect dominates, and the information stored on the even horizon of PBH slows down the evaporation rate by the negative $k$-th power of PBH entropy.

We calculate the graviton flux from PBH evaporation by the public program Blackhawk v2.3 \cite{Arbey:2019mbc,Arbey:2021mbl} and include the burden effects into BlackHawk by modifying the BlackHawk routine $\mathtt{loss\_rate\_M}$ in the file $\mathtt{evolution.c}$~\cite{Arbey:2019mbc,Arbey:2021mbl}.
We add the parameter $\mathtt{q}$, $\mathtt{k}$, and $\mathtt{initial\_mass}$ into this routine and insert a if-else statement such that when $M_{\rm PBH}(t)>qM_{\rm PBH}|_{T_\phi}$, it returns the semiclassical mass loss rate, or else the loss rate is divided by the entropy to the power of $\mathtt{k}$. When a PBH is in the burden phase, its primary emission rate is reduced by the same entropy factor
    \begin{equation}
    \frac{dN_i}{dtdE}(t>t_q)=\frac{1}{S(M_{\rm PBH})^k}\frac{dN_i}{dtdE}.
    \end{equation}
Here, $dN_i/dtdE$ represents the Hawking spectrum predicted from the semiclassical approach. 
Thus we repeat the same modification to the routine $\mathtt{instantaneous\_primary\_}$\\$\mathtt{spectrum}$ in the file $\mathtt{primary.c}$.

The lifetime of PBH is extended when the burden effect is considered. 
\begin{figure}
	\centering
	\includegraphics[width=0.7\linewidth]{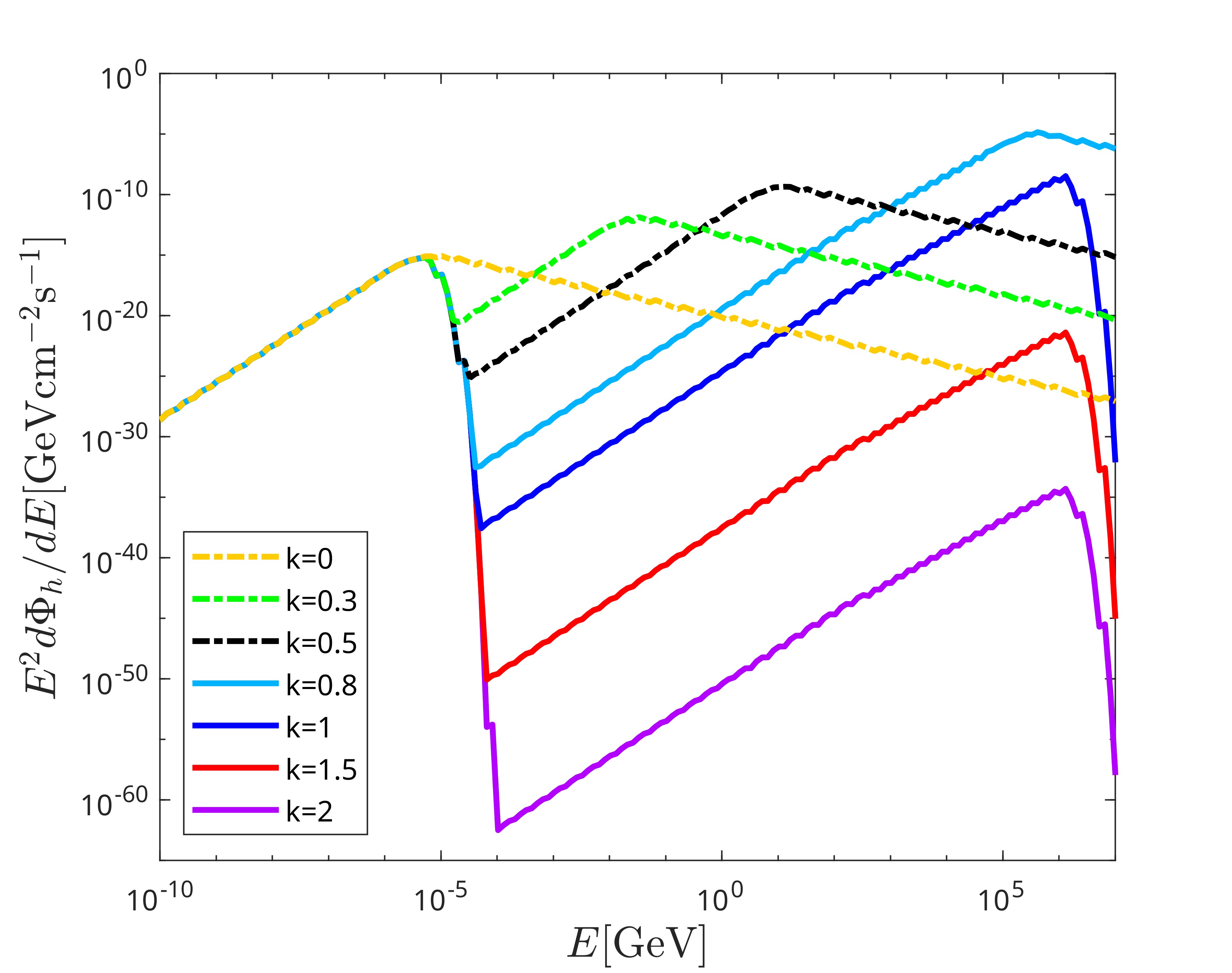}
	\caption{The graviton spectra, observed at present, from evaporation of PBH with $M_{\rm PBH}|_{T_\phi}=10^8\,{\rm g}$, $f_{\rm PBH}|_{T_\phi}=10^{-5}$, and various values of $k$. 
    The first peak around $\mathcal{O}(10^{-5})$ GeV is emitted from PBH in semiclassical phase, while the second peak at higher energy is induced by PBH in burden phase.
    }
	\label{h_spectrum}
\end{figure}
Fig.~\ref{h_spectrum} shows the graviton spectrum of a PBH with mass $M_{\rm PBH}=10^8\,{\rm g}$ at different $k$ values. For $k<0.8$, the burden effect isn't significant enough to sustain the PBH lifetime, and the PBH evaporates before today. Without the burden effect ($k=0$) the spectrum exhibits only one peak around $\mathcal{O}(10^{-5})$ GeV due to redshift. Once the burden effect is included, the mass loss rate suddenly drops when the PBH mass decreases to half of its initial value. The low energy region, which corresponds to large redshift $z$ (small time $t$), is unaffected by the burden effect. In the high energy region, the spectrum drops first and increases subsequently as the PBH evaporates, and the second peak is shifted to higher energy as we increase $k$. For 
$k\geq 1$, the mass loss rate is severely suppressed such that the PBH lifetime is greater than the age of universe. 
The second flux peak is estimated at 
    \begin{equation}\label{secondpeakenergy}
    E\sim T_{\rm PBH}= 5.3\,{\rm MeV}\times\left(\frac{10^{-18}M_\odot}{qM_{\rm PBH}|_{T_\phi}}\right)\approx 2\times10^{5}\,{\rm GeV}.
    \end{equation}
Since the double-peak spectrum is a common feature due to the burden effect, in this work we denote the first and second peak, respectively, corresponding to lower and higher energy in the PBH evaporation spectrum observed today.

\begin{figure}
	\centering
    \includegraphics[width=0.7\linewidth]{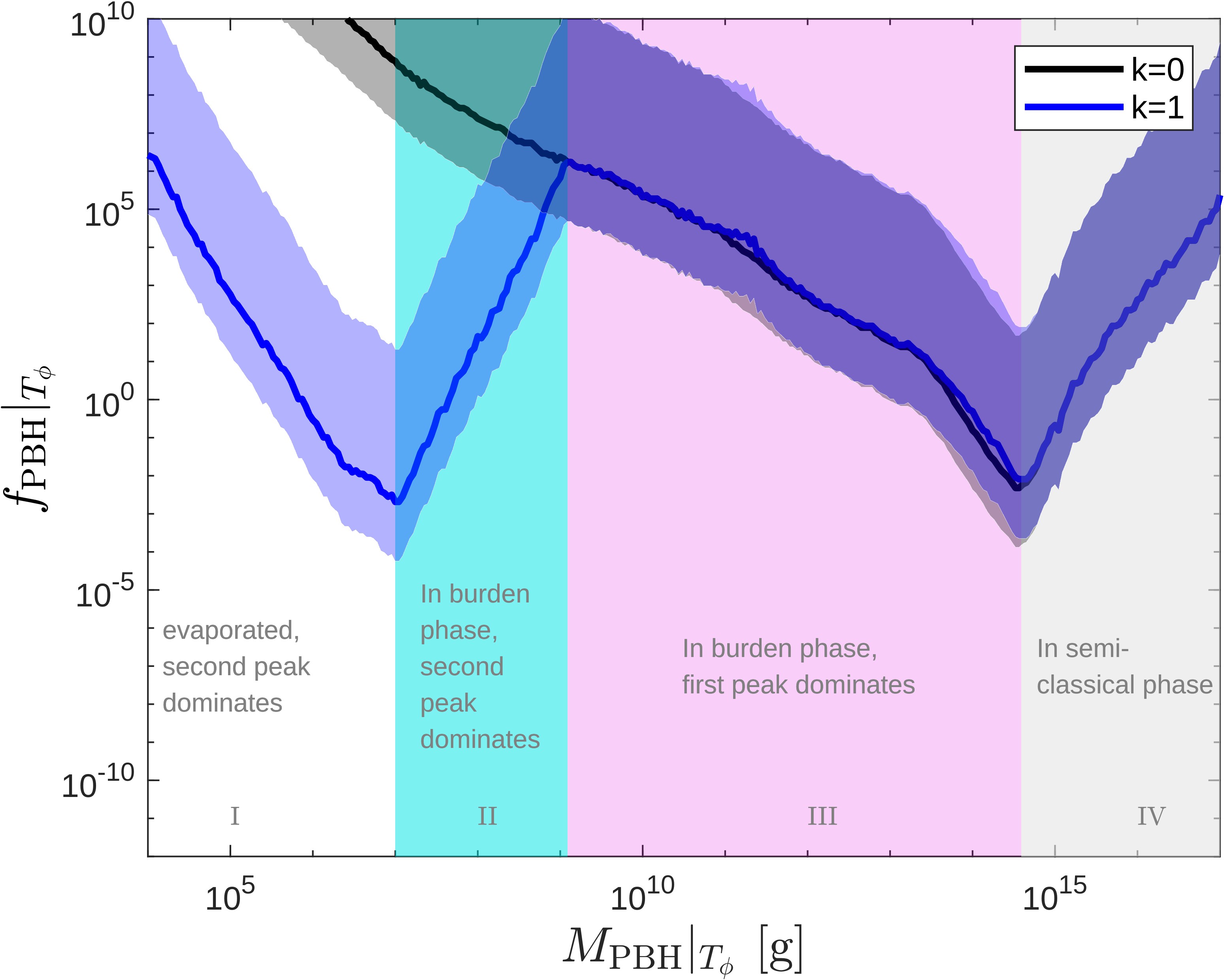}
	\caption{The constraints on $f_{\rm PBH}\vert_{T_\phi}$ for burden parameter $k=0$ and 1. The $f_{\rm PBH}\vert_{T_\phi}$ values above the solid curves are excluded, based on the graviton-photon conversion in the cosmological filaments with optimistic $B_0/{\rm nG}=100$ and sensitivities of extragalactic gamma-ray observations. For $k=1$, it is divided into four regions by the phase of PBH evaporation and first/second peak dominates the gamma-ray observations. Here, we define the PBH mass into {\bf Region-I} to {\bf-IV} for $k=1$. The bands on each constraint illustrate the uncertainty from $B_0$ varying in the range $1\leq B_0/{\rm nG}\leq 600$.}
	\label{fpbh_color}
\end{figure}

\bigskip

\section{Graviton-Photon Conversion}
\label{sec:graviton_photon}

\subsection{Gertsenshtein Effect}

The Gertsenshtein effect describes the phenomenon that a graviton converts to a photon in the presence of a uniform magnetic field~\cite{Gertsenchtein:1962,Boccaletti:1970pxw,Domcke:2020yzq}.
The probability of graviton-photon conversion, which carries the incident energy $\omega$, over a distance $\ell$ and under the angle-averaged transverse magnetic field $B$, can be expressed by
\begin{equation}
    P_{h\to \gamma}= |K_{\rm osc}|^2 \ell^2_{\rm osc} \sin^2\left(\frac{\ell}{\ell_{\rm osc}} \right)\,.
\end{equation}
The oscillation amplitude and characteristic oscillation length are given as
\begin{equation}
|K_{\rm osc}|=\frac{\sqrt{\mu}\kappa B}{1+\mu}\,,~~ \ell_{\rm osc}=\frac{2}{\sqrt{\omega^2(1-\mu)^2+\kappa^2 B^2}}\,,
\end{equation}
where $\kappa^2=16 \pi G_N $ and $\mu$ is the refractive index .
The refractive index relates to the characteristic plasma and the Euler-Heisenberg frequencies, i.e. $\mu=\sqrt{1-(\omega^2_{\rm pl}-\omega^2_{\rm EH})/\omega^2}$~\cite{Dunsky:2025pvd}.
This work focuses on graviton-photon conversion in the cosmological filament.
Since the plasma frequency with the electron number density in the cosmological filament is $\omega_{\rm pl}\sim 10^{-14}$ eV and $\omega_{\rm EH}/\omega \sim 10^{-23} (B/60\,{\rm nG})$, we adopt the approximation $\mu \simeq 1$. The $\ell_{\rm osc}$ becomes almost independent on $\omega$ and vastly exceeds the Hubble radius. As a result, the conversion probability in each filament can be approximated as $P_{h \to \gamma} \simeq 4 \pi G_N \vert B \ell \vert^2$. 

\subsection{Cosmological Filaments}

The gravitons emitted from the PBH can convert to photons in the of uniform magnetic field produced by cosmic filaments \cite{Dunsky:2025pvd}.
The magnetic field from cosmological filaments is given by
    \begin{equation}
    B(z)=B_0(1+z)^2.
    \end{equation}
The observed constraint on $B_0$ is $1\,{\rm nG}-600\,{\rm nG}$. We take the optimistic field $B_0=100\,{\rm nG}$ in our calculation. According to the discussion in earlier section, in a single filament, the conversion probability is approximated as~\cite{Dunsky:2025pvd}
    \begin{equation}
    \label{eq:pz}
    P^{(1)}(z)\approx 4\pi G_N[B(z)l_f(z)]^2,
    \end{equation}
$l_f$ is the traversed length of graviton within the filament, and the fiducial average length is 
    \begin{equation}
    l_f=\frac{4}{1+z}\,{\rm Mpc}.
    \end{equation}
The number of filaments over a distance $dr$ is given by
    \begin{equation}
    dN_f=f_{\rm vol}(z)\frac{dr}{l_f(z)}=\frac{f_{\rm vol}(z)}{l_f(z)}\frac{dz}{H(z)(1+z)},
    \end{equation}
Assuming graviton conversions occur mainly at low redshift, then $f_{\rm vol}(z)\approx0.15$. 
Then the probability for graviton conversion and the differential conversion probability are
    \begin{subequations}
    \begin{align}\label{probability}
    P(z)&=\int dN_f P^{(1)}(z)\,,\\
    \frac{dP}{dz}&=\frac{f_{\rm vol}(z)}{l_f(z)}\frac{P(z)}{H(z)(1+z)}\,,
    \label{dp_dz}
    \end{align}
    \end{subequations}
respectively.
The graviton flux from the PBH evaporation at the redshift $z_c$ is 
    \begin{equation}\label{dphih_dE}
    \frac{d\Phi_h}{dE}(z_c)=\frac{f_{\rm PBH}|_{T_\phi}\Omega_{\rm DM}\rho_c(t_0)}{M_{\rm PBH}|_{T_\phi}}\int^\infty_{z_c}\frac{dz}{H(z)}\frac{dN_h}{d\tilde{E}dt}\Bigg|_{\tilde{E}=E[1+z(t)]},
    \end{equation} 
where $\Omega_{\rm DM}=0.24$ is the dark matter fraction, $\rho_c$ is the critical density, $t_0$ is the age of universe.
$f_{\rm PBH}|_{T_\phi}$ is the PBH fraction at formation time, related to the present mass and fraction by~\cite{Thoss:2024hsr}
    \begin{equation}
    \frac{f_{\rm PBH}|_{T_\phi}}{M_{\rm PBH}|_{T_\phi}}=\frac{f_{\rm PBH0}}{M_{\rm PBH}(t_0)}.
    \end{equation}
The Hubble rate can be parameterized as $H(z)=H_0\sqrt{\Omega_m(1+z)^3+\Omega_r(1+z)^4+\Omega_\Lambda}$
with $H_0=67.36\,{\rm km\  s^{-1} \ Mpc^{-1}}$, $\Omega_m=0.31$, $\Omega_r=9.2\times10^{-5}$, $\Omega_\Lambda=0.68$ \cite{Planck:2018vyg}.
The redshift is the function of time 
    \begin{equation}\label{redshift}
    1+z(t)=\begin{cases}
    (t_0/t)^{2/3}, t>t_{eq}\\
    Ct^{-1/2}, t<t_{eq}
    \end{cases}
    \end{equation}
where $t_{eq}\approx 50000\,{\rm yrs}$ is the matter-radiation equality. $C$ is determined by the continuation of redshift at $t=t_{eq}$. 
Convoluting Eq.(\ref{dp_dz}) and Eq.(\ref{dphih_dE}), we get the total photon flux at redshift $z$
    \begin{equation}
    \frac{d\Phi_\gamma}{dE}(z)=\int^\infty_z dz_c\frac{dP(z_c)}{dz_c}\frac{d\Phi_h}{dE}(z_c).
    \end{equation}

The above formalism adopts the coherence length of the magnetic field in filaments is about 1-10 Mpc~\cite{Amaral:2021mly,Dunsky:2025pvd}, which is comparable to the size of a filament. On the other hand, the typical value of oscillation length $\ell_{\rm osc}\sim 10^4$ Gpc is much larger than the coherent length, and this justifies the approximation Eq.(\ref{eq:pz}) for a single filament.
In the case of the shorter coherence length, for example, it is broken down into $\tilde{n}$ segments, then a factor $1/\tilde{n}$ weaker conversion probability for a single filament would be obtained. In contrast, if the coherence length is across $\tilde{n}$ filaments, the overall conversion probability is enhanced by factor $\tilde{n}$.

\bigskip

\section{PBH Merger}
\label{sec:pbh_merger}
For two burdened PBH in a binary configuration, they may undergo a merging process and reproduce a ``new'' PBH. The new PBH mass is twice of the original burdened mass. Since the Schwarzschild radius $r\propto M_{\rm PBH}$, the horizon area is four times larger and provides considerable memory storage capacity of the initial information. Thus the new PBH emits semiclassically until it reaches half of its mass again.
Considering the merger of monochromatic binary PBH, the differential merger rate is given by \cite{Zantedeschi:2024ram}
    \begin{equation}\label{merging_rate}
    R_{\rm PBH}=\frac{5.7\times 10^{-66}}{\rm cm^3\ s}f_{\rm PBH0}^{\frac{53}{37}}\left(\frac{t_0}{t}\right)^{\frac{34}{37}}\left(\frac{2M_{\rm PBH0}}{10^{10}\,{\rm g}}\right)^{-\frac{32}{37}}S_1S_2,
    \end{equation}
where the suppression factor $S_1\approx0.24$, $S_2(x)\approx {\rm min}[1,9.6\times10^{-3}x^{-0.65}\exp{(0.03\ln^2x)}]$ with $x\equiv (t/t_0)^{0.44}f_{\rm PBH0}$. The extragalactic flux contribution from the merged PBH is 
    \begin{equation}
    \frac{d\Phi_i}{dE}=t_q\int_0^{z_{\rm max}}\frac{dz}{H(z)}R_{\rm PBH}(t(z))\left(\frac{dN_i}{d\tilde{E}dt}\Bigg|_{\tilde{E}=E[1+z(t)]}\right)_{\rm avg}.
    \end{equation}
$(dN_i/d\tilde{E}dt)_{\rm avg}$ is the averaged emission rate of the merged PBH over the semiclassical phase. The upper limit of redshift integration is set to $z_{\rm max}=10$\footnote{We have extended $z_{\rm max}=100$ to check the stability, and no significant modifications of the extragalactic flux has been found.}.

Note that Eq.(\ref{merging_rate})
was not originally derived for the light PBH mass region, which we are interested in. There are additional effects, such as late-time dynamical capture and dynamical captures induced by gravitational wave, but these are subdominant according to Ref.~\cite{Weng:2022ehq}. However, it is known that local non-Gaussianities of primordial curvature perturbation can modify the initial distribution of PBHs, clustering directly at the formation time, potentially enhancing the merger rate by $\mathcal{O}(10^7)$~\cite{Weng:2022ehq,DeLuca:2021hde}. We used Eq.(\ref{merging_rate}) to obtain the conservative merge rate.

\bigskip

\section{Constraints on PBH}
\label{sec:constraint}

We compare the photon flux from graviton conversion or merged PBHs with the current 
JEM-X/IBIS-ISGRI/FermiLAT extragalactic gamma-ray upper limits and future e-ASTROGAM/CTA South/LHAASO/HiSCORE sensitivities. The upper bounds of $f_{\rm PBH}|_{T_\phi}$ can be derived.

In Fig. \ref{fpbh_color}, we show the constraint on $f_{\rm PBH}|_{T_\phi}$ with $k=0.0$ and $1.0$ according to the graviton-photon conversion. The plot is divided into four regions to indicate four different photon flux behaviors of $k=1.0$. The PBH in {\bf Region-I} ($M_{\rm PBH}|_{T_\phi} \leq 10^7$ g), although prolonged by memory burden, is completely evaporated today. The evaporation gets stronger, because the increase of PBH mass brings its lifetime closer to the age of universe, and eventually the flux gets enhanced due to the final evaporation. Hence, the minimum of $f_{\rm PBH}|_{T_\phi}\simeq 5\times 10^{-3}$ occurs at $M_{\rm PBH}|_{T_\phi} \simeq 10^7$ g. In fact, the second peak of photon flux is constrained by the high-energy CTA South/LHAASO/HiSCORE observations.
For $10^{7}\,{\rm g} \leq M_{\rm PBH}|_{T_\phi}\leq 10^{9}\,{\rm g}$ ({\bf Region-II}), PBH enters the burden phase at present Universe, and its second peak in graviton spectrum is suppressed by entropy increasing with $M_{\rm PBH}|_{T_\phi}$, 
as well as the upper limit of $f_{\rm PBH}|_{T_\phi}$ tends to increase with $M_{\rm PBH}|_{T_\phi}$. In {\bf Region-III}, the first peak amplitude in graviton spectrum overtakes the second peak and is constrained by the low-energy JEM-X/IBIS-ISGRI/e-ASTROGAM observations. Thus, $f_{\rm PBH}|_{T_\phi}$ decreases again. The PBH masses in {\bf Region-IV} ($M_{\rm PBH}|_{T_\phi} \gtrsim4\times10^{14}\,{\rm g}$) are too large 
to evaporate half of the initial mass at present. They remain in semiclassical phase today and produce a single-peak photon flux. 
The constraint on $f_{\rm PBH}|_{T_\phi}$ becomes weaker, as the mass loss rate is suppressed for heavier PBH. 
For most of the mass region, the upper limit of $f_{\rm PBH}|_{T_\phi}> 1$ is allowed, except for those within the ranges of $7.5\times10^5\ {\rm g}\leq M_{\rm PBH}|_{T_\phi}\leq 4.4\times10^7\ {\rm g}$ and $7.7\times10^{13}\ {\rm g}\leq M_{\rm PBH}|_{T_\phi}\leq 1.4\times10^{15}\ {\rm g}$

\begin{figure}
	\centering
    \includegraphics[width=0.496\linewidth]{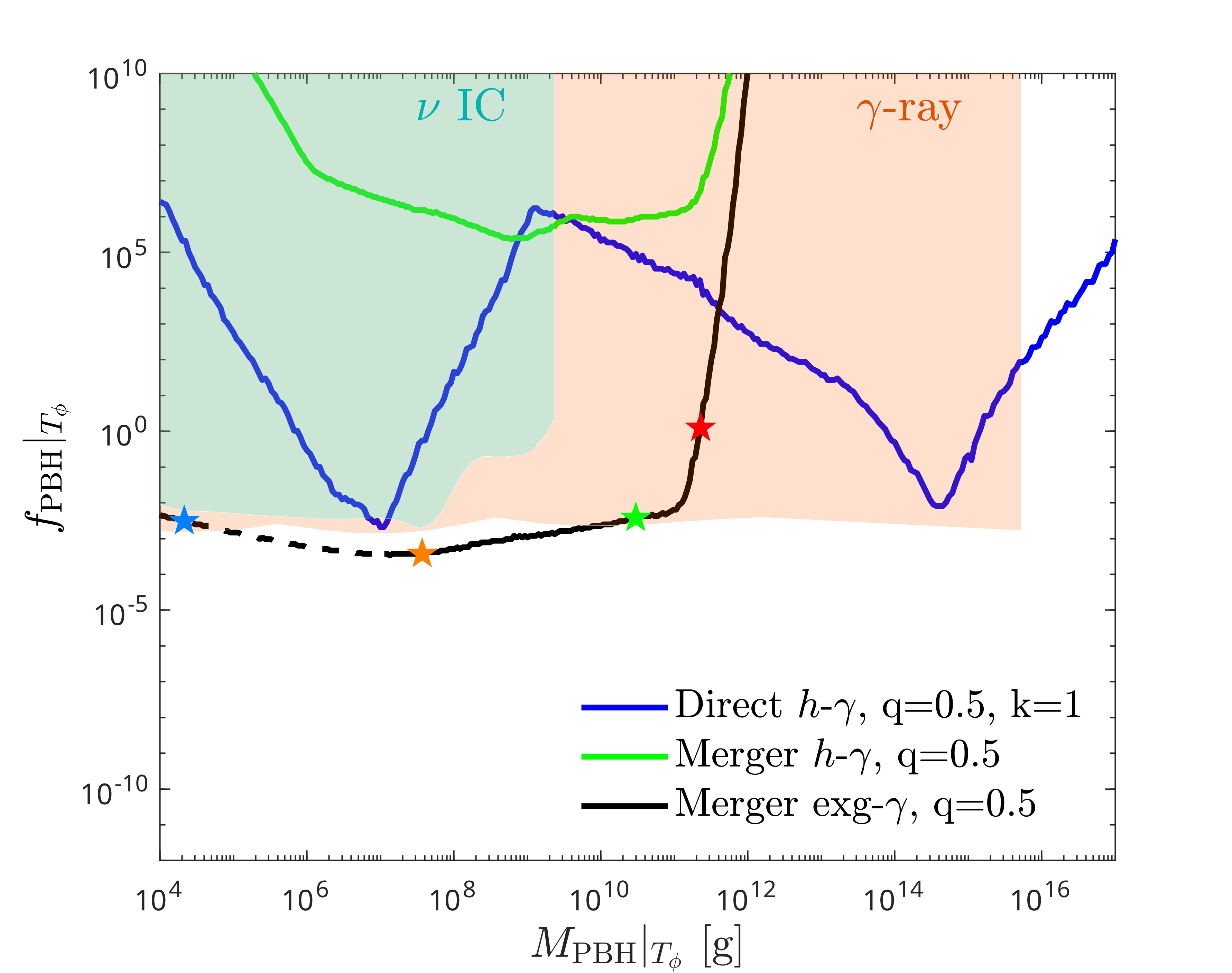}
    \includegraphics[width=0.496\linewidth]{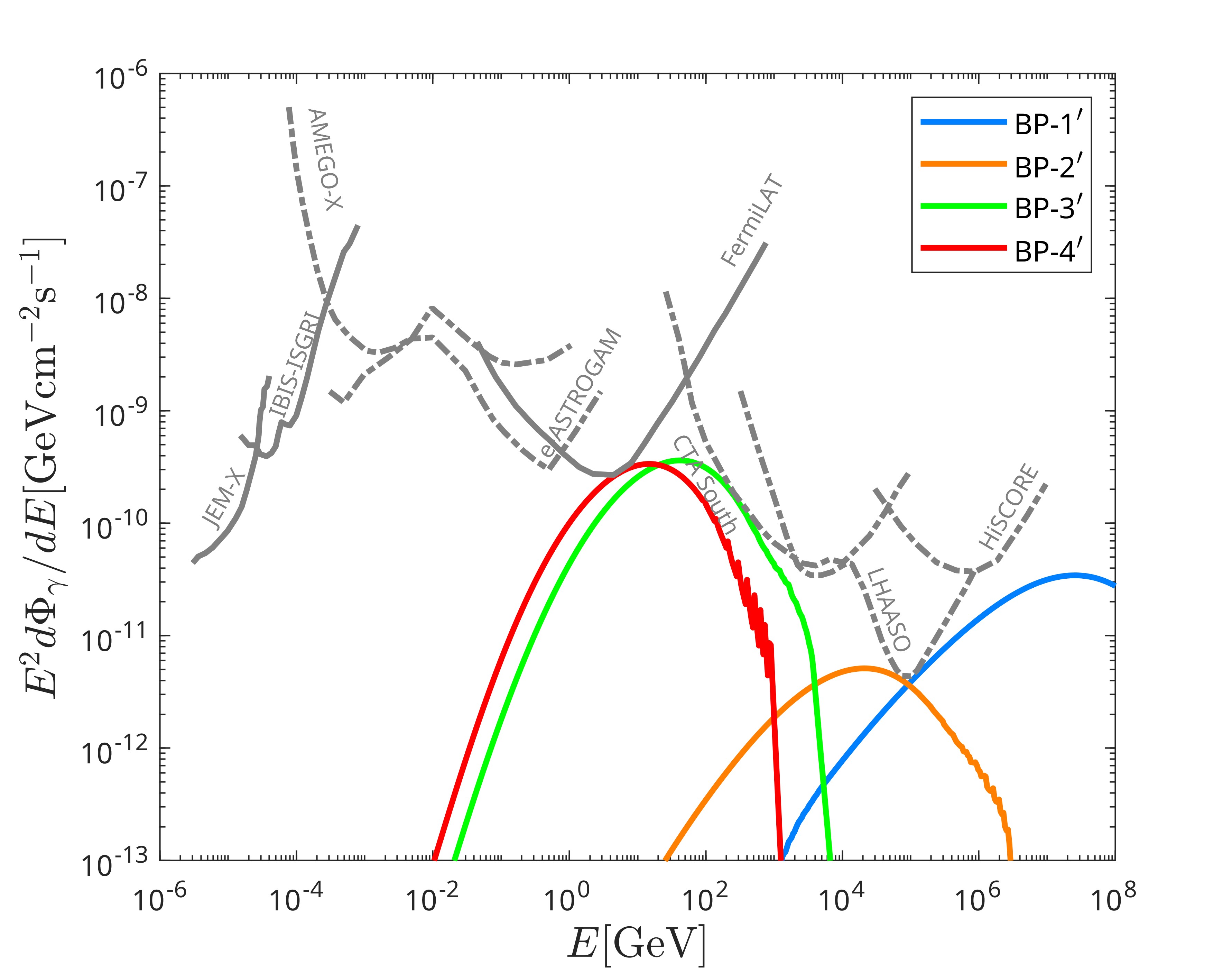}
	\caption{Left panel:
    The comparison of $f_{\rm PBH}|_{T_\phi}$ constraints via photon flux from graviton-photon conversion (blue curve), PBH merger (black curve), and combination of the two (green curve).
    Compared with other analysis, the cyan-shaded region is the merger neutrino constraint derived from IceCube Collaboration \cite{Zantedeschi:2024ram}, and the orange-shaded region is the combined gamma-ray constraint (galactic and extragalactic) from \cite{Dondarini:2025ktz}. 
    The dashed part of the ``Merger exg-$\gamma$" indicates the evaporated mass before present epoch for $k=1$, and the "$\star$" label the benchmark points ({\bf BP}$^\prime$s) with corresponding values in Table~\ref{table_mer}. Right panel: The corresponding photon spectra of {\bf BP}$^\prime$s from left panel, and their values are listed in Table~\ref{table_mer}
	\label{compare}}
    \end{figure}

The comparison of the constraints from the direct burdened PBH emission and the PBH merger is shown in the left panel of Fig.~\ref{compare}\footnote{Caveat: The semiclassical PBH formed after merging of two memory-burdened PBHs is based on the arguments that the newly formed BH extends four times of the area than the combined areas of the two initial BHs, increasing memory-storage capacity~\cite{Zantedeschi:2024ram}. Ref.~\cite{Dvali:2025sog}, however, has an opposite statement. This phenomenon has not been confirmed by rigorous theory or model, therefore our limit from the merging PBH should be taken with caution.}. The spectrum of the merger depends only on $q$, but is mildly dependent on $k$. As we decrease the $q$ value, the merged PBH stays longer in the semiclassical phase and enhances the total flux. However, for comparison with bounds from graviton-photon conversion and others, 
we fix $q=0.5$. We computed the extragalactic photon flux from the merger; the upper bound of $f_{\rm PBH}|_{T_\phi}$ is shown by the black solid curve in the left panel of Fig.~\ref{compare} (The black dashed part indicates the evaporated PBH mass region for $k=1$ before present epoch). For $M_{\rm PBH}|_{T_\phi}\lesssim 10^{11}\,{\rm g}$, 
the significant evaporation needs to be compensated by $f_{\rm PBH}|_{T_\phi}$ of the order $\mathcal{O}(10^{-3})$,
but for $M_{\rm PBH}|_{T_\phi}\gtrsim 10^{11}\,{\rm g}$ the hawking spectrum in semiclassical phase 
decreases with increasing mass
so the value of $f_{\rm PBH}|_{T_\phi}$ grows rapidly. This upper limit of PBH merging is insensitive with the parameter $k$, at least it provided that the value of $k$ is large enough to ensure that the PBHs survive until the redshift $z_{\rm max}$. The black-dashed curve in left panel of Fig.~\ref{compare} indicates the PBHs evaporated before $z_{\rm max}$ for $k=1$.
The photon fluxes of the four benchmark points on the merger extragalactic photon constraint are shown in the right panel of Fig.~\ref{compare}.
In addition, we incorporate graviton-photon conversion into the PBH merging scenario and then obtain the bound of $f_{\rm PBH}|_{T_\phi}$ in the green curve in the left panel of Fig.~\ref{compare}.
However, this is much weaker than the graviton-photon conversion limit, except for the tiny mass range $7.6\times10^{8}\ {\rm g}\leq M_{\rm PBH}|_{T_\phi}\leq3.2\times10^{9}\ {\rm g}$. 

	\begin{table}
	\footnotesize
	\centering
	\begin{tabular}{c| c c c c}
		\hline
        &{\bf BP-1$^\prime$}&{\bf BP-2$^\prime$}&{\bf BP-3$^\prime$}&{\bf BP-4$^\prime$}\\
        \hline
        $M_{\rm PBH}|_{T_\phi}/{\rm g}$&$2.14\times10^{4}$&$3.70\times10^{7}$&$2.98\times10^{10}$&$2.28\times10^{11}$\\
        $f_{\rm PBH}|_{T_\phi}$&$3.08\times10^{-3}$&$3.74\times10^{-4}$&$3.81\times10^{-3}$&$1.25$\\
        \hline
	\end{tabular}
	\caption{The $(M_{\rm PBH}\vert_{T_\phi},f_{\rm PBH}\vert_{T_\phi})$ for benchmark points from Fig.~\ref{compare}.}
	\label{table_mer}
\end{table}

\begin{figure}
	\centering
    \includegraphics[width=0.496\linewidth]{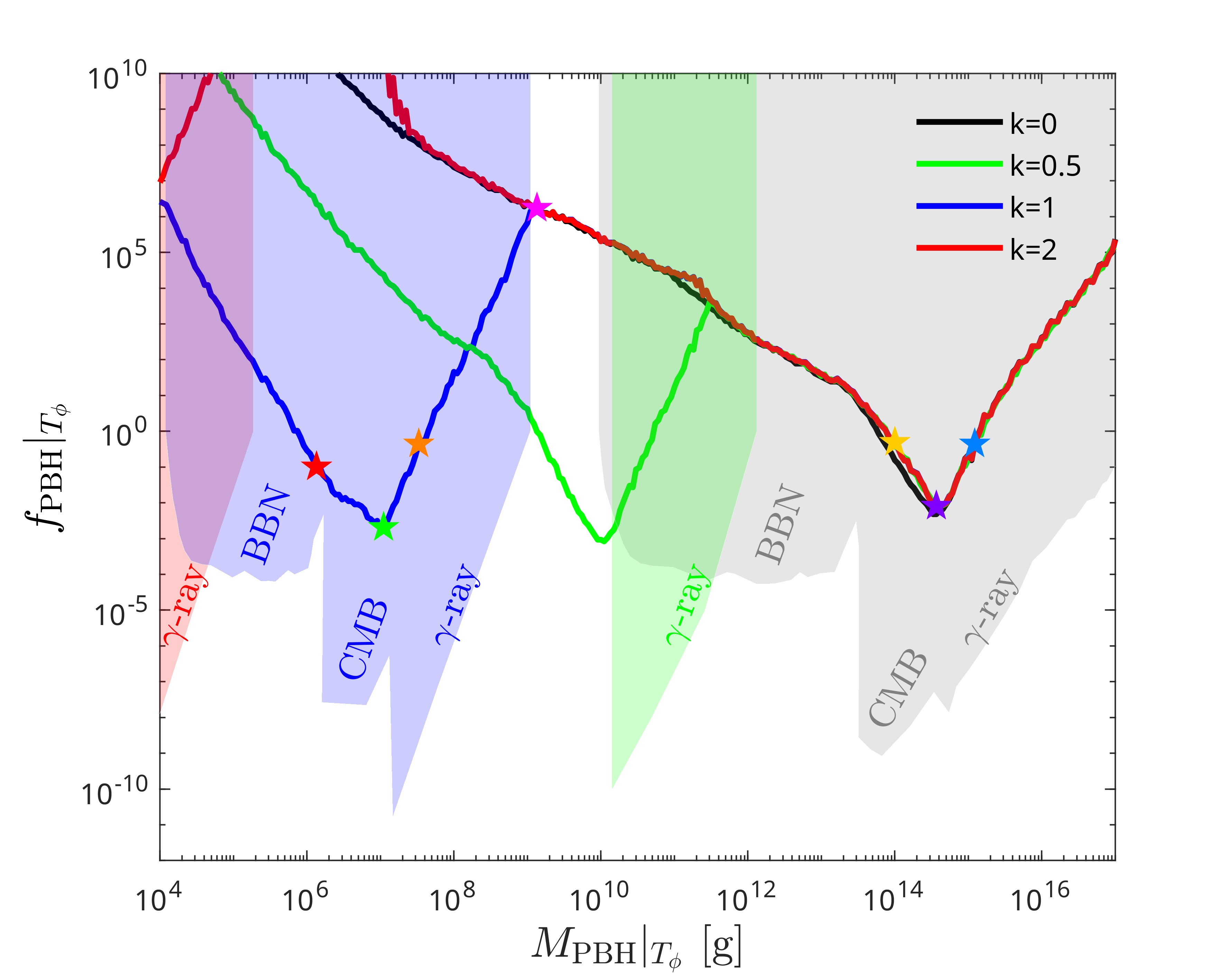}
    \includegraphics[width=0.496\linewidth]{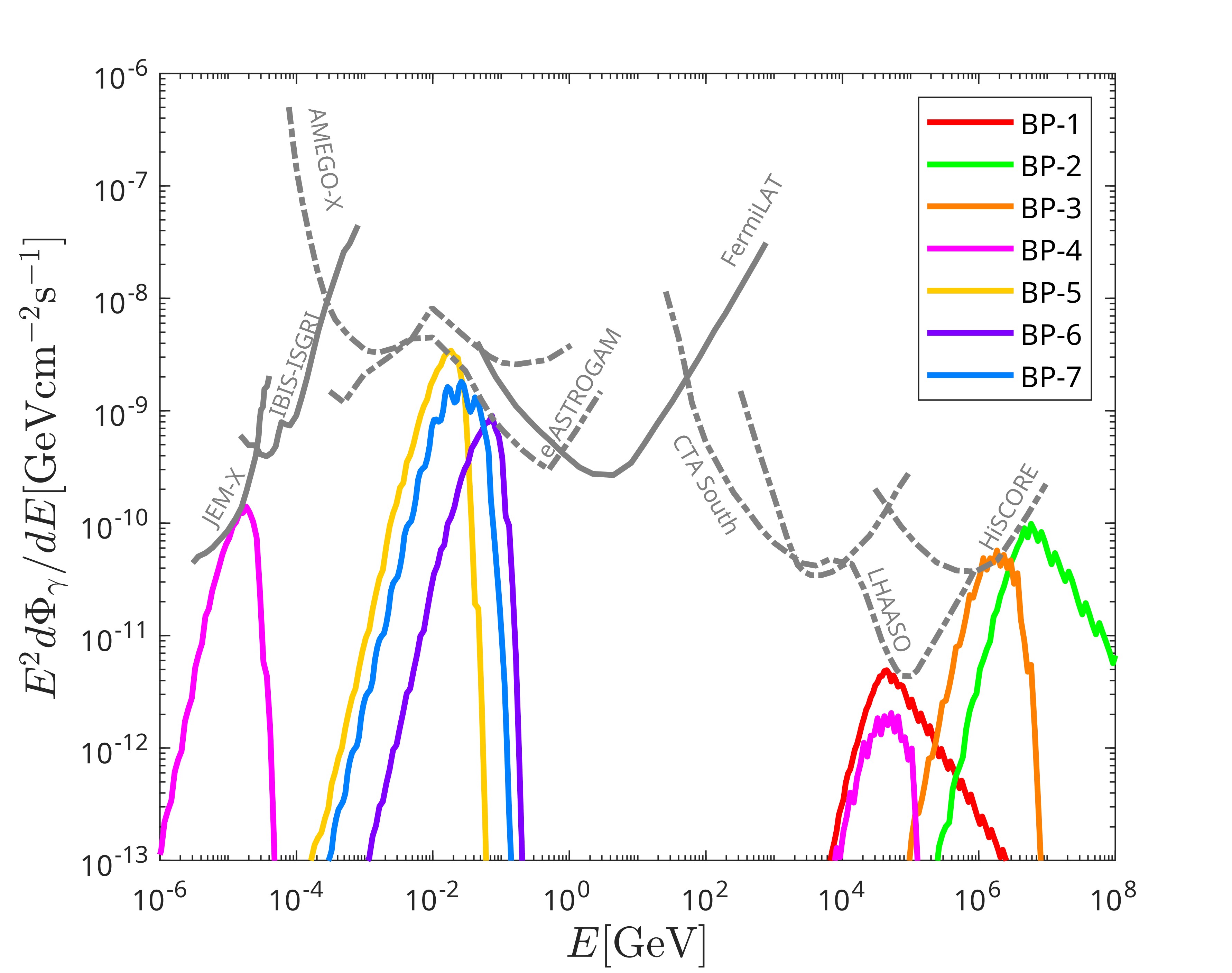}
	\caption{Left panel: The solid curves are constraints on $f_{\rm PBH}|_{T_\phi}$ from graviton-photon conversion for $k=0\,,0.5\,,1.0\,,2.0$. The {\bf BP}s, labeled by ``$\star$'', are selected along the $k=1.0$ sensitivity curve. Their parameters $(M_{\rm PBH}\vert_{T_\phi},f_{\rm PBH}\vert_{T_\phi})$ are shown in Table~\ref{table1}.
    Compared with other analysis, the red-, blue-, and green-shaded regions are respectively excluded for $k=2.0$, $k=1.0$~\cite{Chaudhuri:2025asm}, and $k=0.5$~\cite{Thoss:2024hsr}; the gray-shaded region is excluded for PBH in semiclassical phase which is applicable for all $k$ value. Right panel: The photon spectra of the corresponding {\bf BP}s in left panel, and their values are listed in Table~\ref{table1}.} 
	\label{fpbh}
\end{figure}

In the left panel of Fig.~\ref{fpbh}, we add the limits for $k=0.5$ and $2.0$. 
As $k$ increases from 0, the burden effect starts to suppress Hawking radiation and extend the lifetime of PBH, and the trends of $f_{\rm PBH}|_{T_\phi}$ (for $0<k\leq1$) are split into four different mass regions as 
shown in Fig.~\ref{fpbh_color}.
The $k=0$ constraint exhibits only {\bf Region-IV} 
since there's no burden effect and the PBH evaporates semiclassically. 
For $k > 0$, once we include the graviton-photon conversion, the constraint for {\bf Region-I} and {\bf Region-II} become more stringent.
The valley between {\bf Region-I} and {\bf -II} is shifted to lower $M_{\rm PBH}|_{T_\phi}$ as the burden effect gets stronger. However, the $f_{\rm PBH}|_{T_\phi}$ for $k=2$ exhibits a severe increment from its {\bf Region-III} to {\bf-II}. This is because the first peak of its photon spectrum is red-shifted to lower energy as $M_{\rm PBH}|_{T_\phi}$ reduces, the PBH with lower mass enters the burden phase earlier than heavier PBH, so the spectrum experiences a longer redshift. As the first peak becomes smaller than the lowest gamma-ray sensitivity limit (JEM-X), the $f_{\rm PBH}|_{T_\phi}$ grows inevitably. The constraints of $f_{\rm PBH}|_{T_\phi}$ for $k=$0.5, 1.0, and 2.0 from Ref.~\cite{Thoss:2024hsr,Chaudhuri:2025asm} are included in Fig.~\ref{fpbh} by shaded regions. They considered photons emitted directly from a PBH, which propagate freely after the cosmic recombination ($\sim378,000$ yrs after the big bang). In their case, the first peak of photon flux, which is produced before the recombination time, is only indirectly constrained via CMB distortion or Big Bang Nucleosynthesis (BBN).
On the other hand, the gravitons interact extremely weakly with the SM particles, and thus their propagation is free streaming even before CMB recombination. Therefore, the first peak of photon flux from converted graviton, which is not suppressed by the burden effect, can 
provide a complementary limit on the abundance of PBH.

In the left panel of Fig.~\ref{fpbh}, seven benchmark points are labeled along the $k=1$ constraint, the corresponding values are listed in Table \ref{table1}. The right panel of Fig. \ref{fpbh} shows the corresponding photon flux of {\bf BP}s. Except for {\bf BP-4}\footnote{Notice {\bf BP-4} is for the purpose to demonstrate the double-peak gamma-ray spectrum under sensitivities in near future, but its relic abundance is not consistent with the cosmological observations, exceeding the matter component.}, all benchmark points are bounded to have $f_{\rm PBH}|_{T_\phi} \leq 1$.
{\bf BP-1, -2}, and {\bf -3} are in the second peak-dominated region, and their second peaks are constrained by high-energy gamma-ray sensitivity CTA South, LHAASO, and HiSCORE. 
From Fig.~\ref{fpbh_color} and Fig.~\ref{fpbh}, we see that {\bf BP-4} locates at the boundary of {\bf Region-II} and {\bf Region-III}, even though $f_{\rm PBH}|_{T_\phi}$ far exceeds unity, its photon spectrum features two comparable peaks from observations. Thus, the transition of the dominant peak can be read from {\bf BP-1} to {\bf BP-7}.
For {\bf BP-5} to {\bf BP-7}, the first peaks dominate, so their gamma-ray spectra are located in the sub-GeV region. The energy of the first peak increases from {\bf BP-4} to {\bf BP-6}. This is because {\bf BP-4} (relative lighter PBH mass) enters the burden phase earlier than {\bf BP-6} (relative heavier PBH mass), so the graviton spectrum of the former experiences a longer redshift. 
For {\bf BP-1} to {\bf BP-3}, their second peaks dominate the observations.
The energy of the second peak decreases from {\bf BP-2} to {\bf BP-4} because of Eq.~(\ref{secondpeakenergy}). 
According to {\bf BP-1} evaporated time $t_{\rm evp}=7.7\times10^{11}\,{\rm sec}$, Eq.~(\ref{secondpeakenergy}), and Eq.~(\ref{redshift}), we can estimate that the peak energy of the graviton flux is red-shifted to 2.6 TeV,
which is about one order lower than the converted photon peak energy $\approx44\,{\rm TeV}$. Similarly, for {\bf BP-7}, we have graviton flux peaked at $\sim 8.7$ MeV,
which is again close to the converted photon peak energy $\approx 26\,{\rm MeV}$.

The above analysis and sensitivities are based on the assumption that all particle species, including gravitons, are universally suppressed by the entropy factor.
However,
in the context of large extra dimension (LED),
the suppression or enhancement of greybody factors depend on particle species.
In order to compute the greybody correction, one needs to sum over the BH emission phase space of individual particle.
Because the emission of SM (Standard Model) particles is confined to three dimensional space, in contrast, gravitons propagate in the bulk with much larger emission phase space.
According to the results of Ref.~\cite{Friedlander:2022ttk,Ettengruber:2025kzw}, the graviton emission power in BH mass loss changes from 0.1\% ($n=0$) to 14.4\% ($n=6$), where $3+n$ represents total spatial dimensions. It is interesting to investigate nonuniversal suppression for future study.

	\begin{table}
	\footnotesize
	\centering
	\begin{tabular}{c| c c c c}
		\hline
        &{\bf BP1}&{\bf BP2}&{\bf BP3}&{\bf BP4}\\
        \hline
        $M_{\rm PBH}|_{T_\phi}/{\rm g}$ &$1.35\times10^{6}$&$1.11\times10^{7}$&$3.32\times10^{7}$&$1.35\times10^{9}$\\        $f_{\rm PBH}|_{T_\phi}$&$1.02\times10^{-1}$&$2.06\times10^{-3}$&$4.45\times10^{-1}$&$1.73\times10^{6}$\\
        \hline
        \hline
        &{\bf BP5}&{\bf BP6}&{\bf BP7}&\\
        \hline
        $M_{\rm PBH}|_{T_\phi}/{\rm g}$ &$1.00\times10^{14}$&$3.67\times10^{14}$&$1.22\times10^{15}$\\$f_{\rm PBH}|_{T_\phi}$&$4.94\times10^{-1}$&$8.12\times10^{-3}$&$4.45\times10^{-1}$\\
        \hline
	\end{tabular}
	\caption{The $(M_{\rm PBH}\vert_{T_\phi},f_{\rm PBH}\vert_{T_\phi})$ for benchmark points from Fig.~\ref{fpbh}.}
	\label{table1}
\end{table}
    


\bigskip

\section{Summary}
\label{sec:summary}
PBHs lighter than $10^{14}$ g have already entered the memory burden phase at the present epoch. However, there are opportunities where we can detect the particles evaporated from their earlier stage, semiclassical PBHs. In this work, we study two scenarios, i) the gravitons emitted from PBHs and ii) the merger of PBHs.

We compute the constraints on the PBH fraction for these two scenarios. The first is the spectrum from the direct Hawking radiation. 
Since the photon spectrum has been extensively studied, we focus on the converted photon flux from the graviton spectrum of PBH.
The gravitons convert to photons by the Gertsenshtein effect, particularly in the cosmological filaments, which possess a uniform magnetic field that depends on the cosmological redshift.  Due to the insensitivity of the graviton to SM particles, the graviton emission from the very beginning of PBH formation could affect the photon spectrum after cosmic recombination, providing the opportunity to probe light PBHs in the semiclassical phase. Thus, the graviton and the photon flux at present would correlate with the mass and fraction of PBHs at the formation time.
The graviton spectrum and thus the converted photon spectrum are characterized by a double-peak structure; the first peak is associated with an earlier stage of PBH in semiclassical phase and is significantly redshifted to lower energy, while the second peak at higher energy is emitted from PBH in burdened phase and thus experienced a mild redshift. However, the latter is suppressed by the black hole entropy and depends on $k$ parameter.
The upper bounds of $f_{\rm PBH}|_{T_\phi}$ are obtained via comparing the converted photon flux with extragalactic gamma-ray observational sensitivities.
As a result, the upper bounds of $f_{\rm PBH}|_{T_\phi}$ exhibit a double trough shape and vary with different PBH masses which are divided into four distinct regions for $k=1$.
Except for the {\bf Region-IV}, the $f_{\rm PBH}|_{T_\phi}$ depends on the strength of the suppression parameter $k$. Based on optimistic magnetic field assumption and simplified oscillation treatment, for $k=0$, the $f_{\rm PBH}|_{T_\phi}>1$ is excluded in the range 
$5.8\times10^{13}\ {\rm g}<M_{\rm PBH}|_{T_\phi}<1.4\times10^{15}\ {\rm g}$.
For $k=0.5$ ($1.0$), there is an additional region 
$1.3\times10^{9}\ {\rm g}<M_{\rm PBH}|_{T_\phi}<5.2\times10^{10}\ {\rm g}$
($7.5\times10^{5}\ {\rm g}<M_{\rm PBH}|_{T_\phi}<4.4\times10^{7}\ {\rm g}$)
where $f_{\rm PBH}|_{T_\phi}>1$ is disfavored.
Compared to Ref.~\cite{Thoss:2024hsr,Chaudhuri:2025asm}, the converted photon constraints are seven to eight orders weaker than the photons from the burden phase, but it provides a complementary probe on PBH in semiclassical phase before CMB.

The second scenario we considered is the PBH merger, where young semiclassical black holes with unsuppressed evaporation are formed at present time.
Theoretically, this scenario is highly model dependent and does not have sufficient theory support.
For the phenomenology study,
the extragalactic photon from merged PBHs, which depends only on the merging rate, has a stronger constraint than the converted photon scenario. The sensitivity of $f_{\rm PBH}|_{T_\phi}$ is restricted between $6.5\times10^{-3}$ and $3.7\times10^{-4}$ for $M_{\rm PBH}|_{T_\phi}\lsim 10^{11}\,{\rm g}$ and exceeds unity for $M_{\rm PBH}|_{T_\phi}\gtrsim 2.2\times10^{11}\,{\rm g}$.
In addition, the constraint of combined graviton-photon conversion and merger scenarios is extremely weak and incompetable with other limits, which suffers from the double suppression of the merging rate Eq.~(\ref{merging_rate}) and the conversion probability Eq.~(\ref{probability}).

\bigskip

\section*{Acknowledgment}

We acknowledge the kind support of the National Science and Technology Council of Taiwan R.O.C., with grant number NSTC 111-2811-M-007-018-MY3.
P.Y.T. acknowledges support from the Physics Division of the National Center for Theoretical Sciences of Taiwan R.O.C. with grant NSTC 114-2124-M-002-003.
Y.M.Y. is supported in part by Grant No. 113J0073I4 and NSTC Grant No. 111B3002I4. and by the doctoral scholarship from the Ministry of Education of Taiwan R.O.C.

\bigskip
\newpage

\appendix
\section{Continuous Burden Phase Transition}
\begin{figure}
    \centering
    \subfloat[$p=20$]{\includegraphics[width=0.47\linewidth]{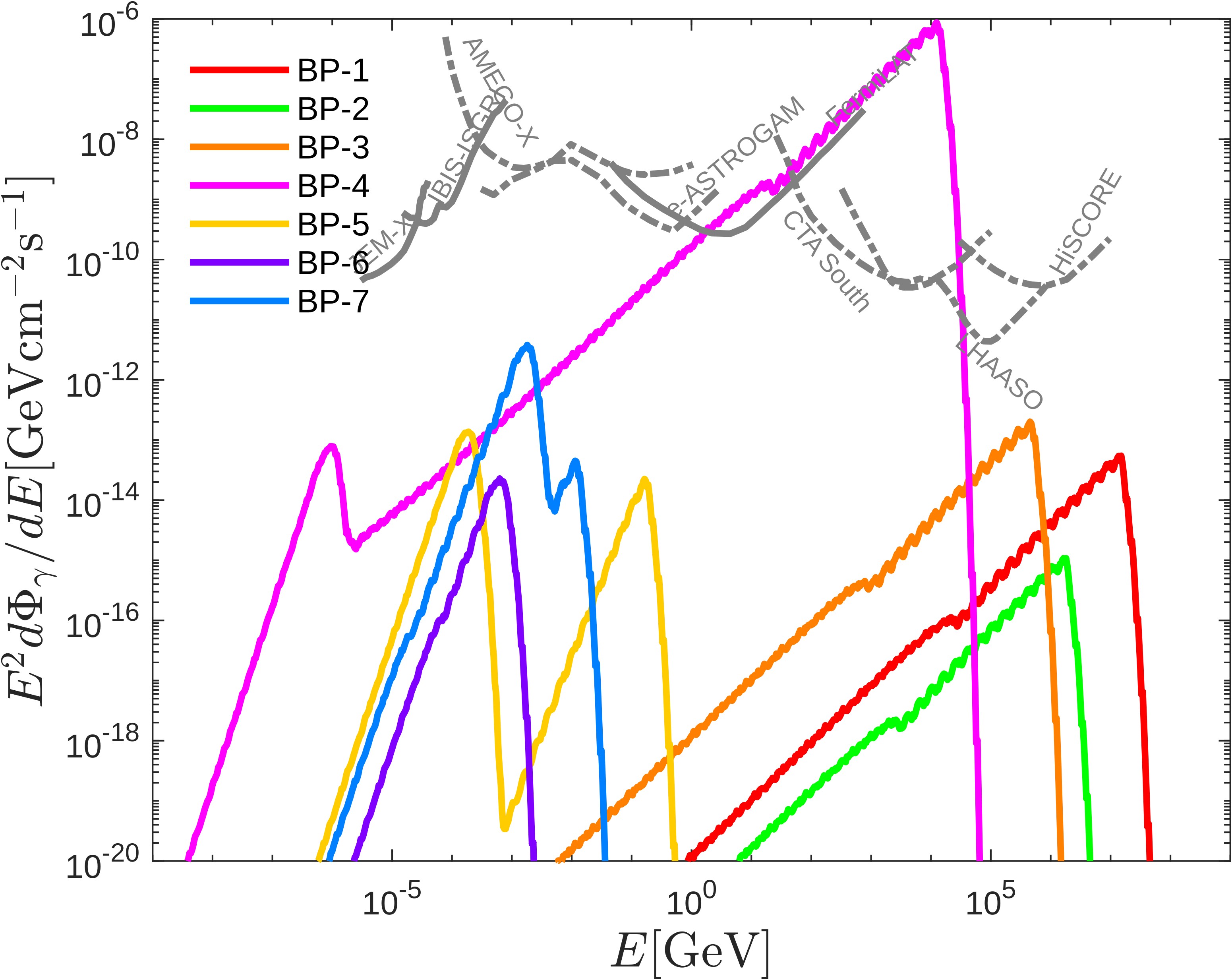}}\quad
    \subfloat[$p=200$]{\includegraphics[width=0.47\linewidth]{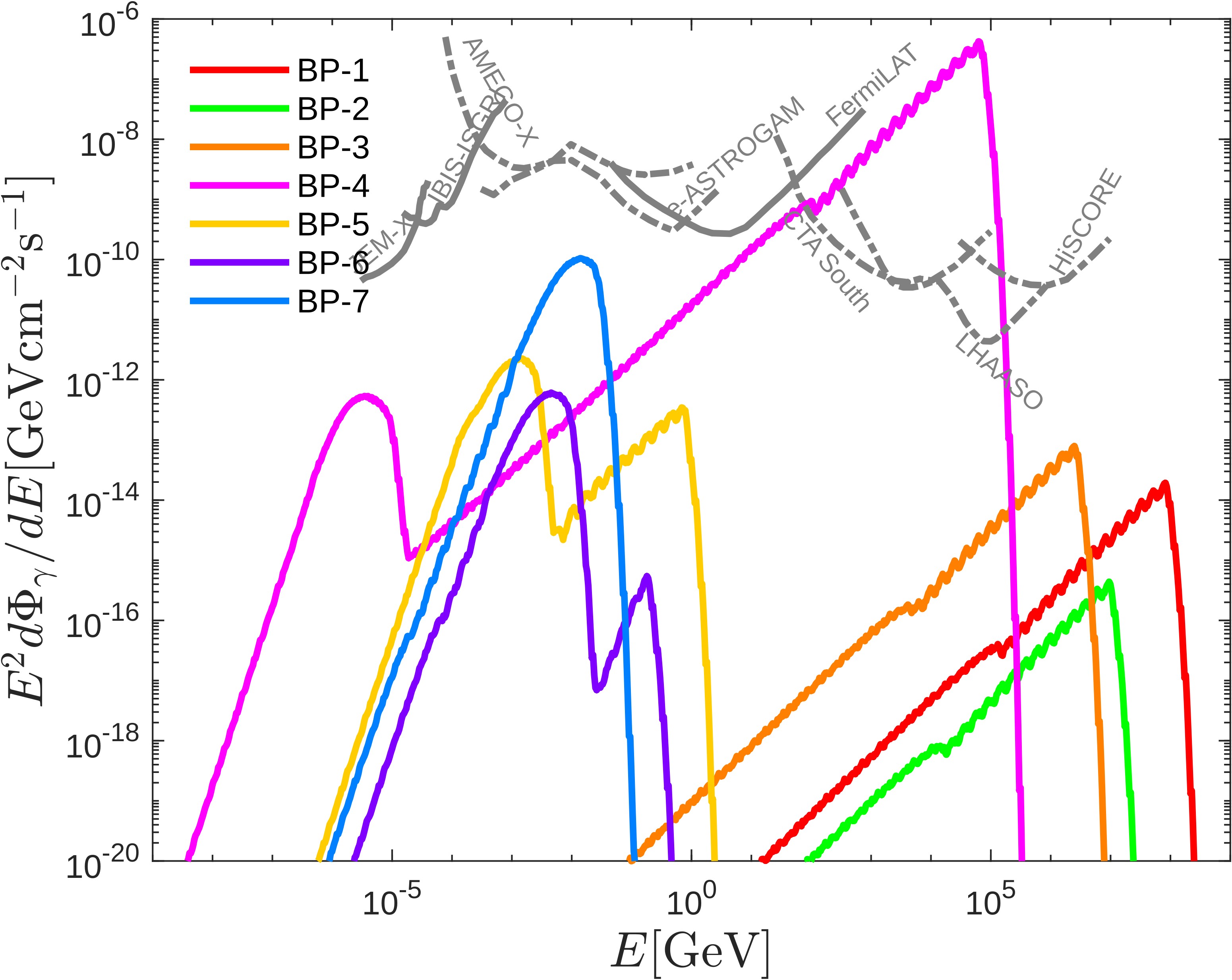}}
    
    \subfloat[$p=2000$]{\includegraphics[width=0.47\linewidth]{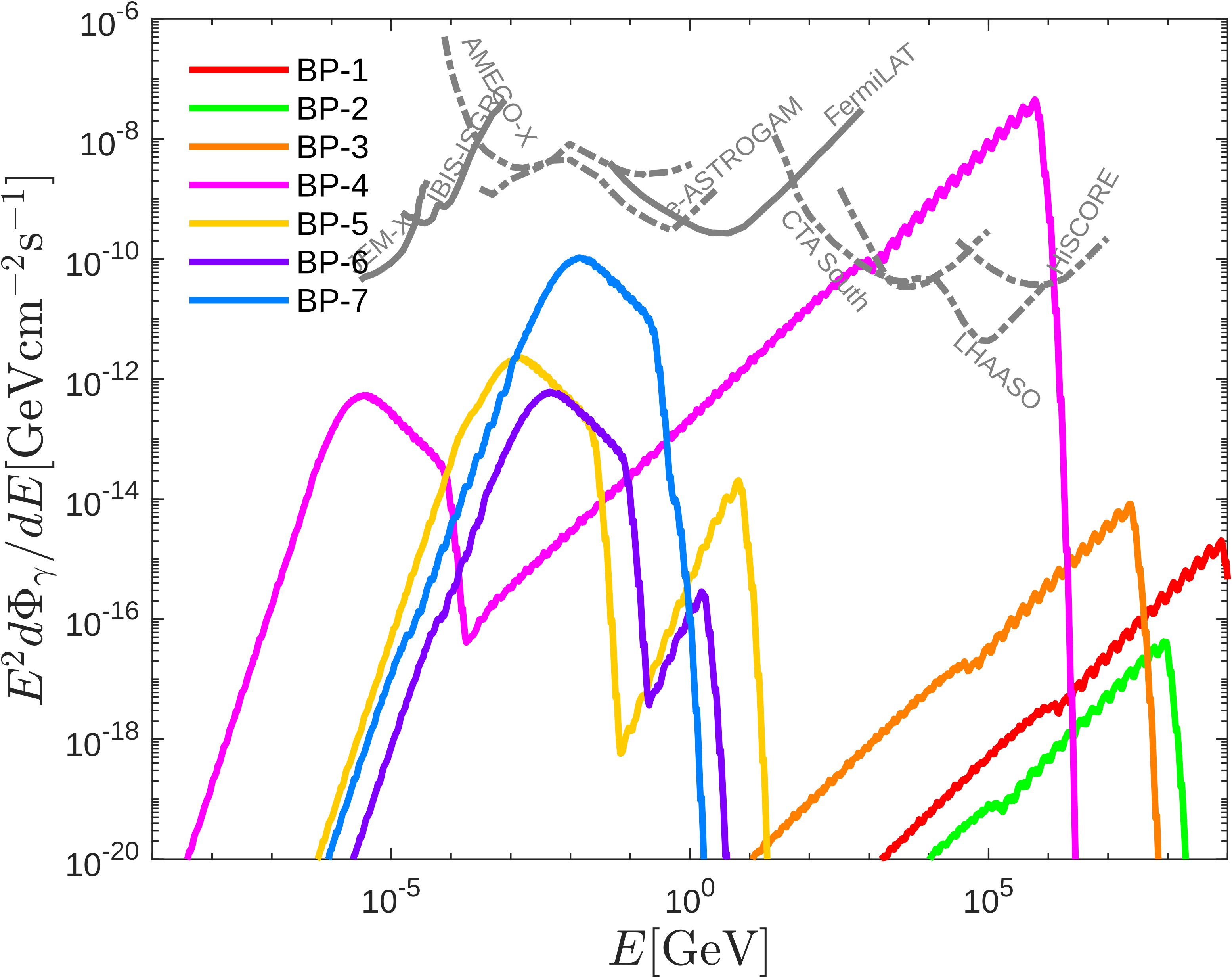}}\quad
    \subfloat[$p=20000$]{\includegraphics[width=0.47\linewidth]{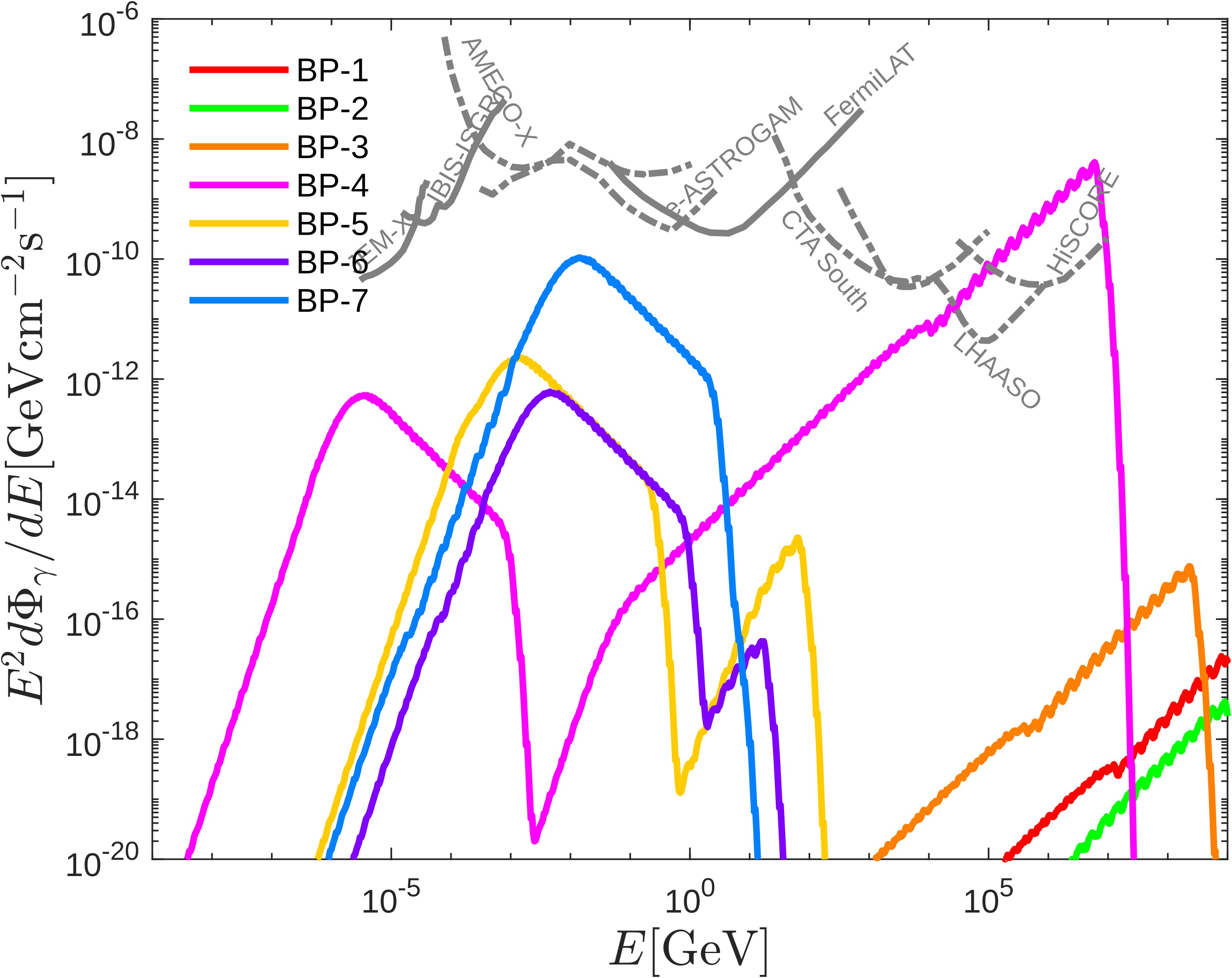}}
    \caption{The photon spectra of the BPs' in Table. \ref{table1}}
    \label{continuous}
\end{figure}

We follow Ref.~\cite{Dvali:2025ktz} to examine how continuous burden-phase transition affects our graviton-photon conversion spectra. The particle emission rate is estimated by
    \begin{equation}
    \Gamma=S^{-\Delta N}\Gamma_{\rm sc}
    \end{equation}
with 
    \begin{equation}
    \Delta N=\frac{p\sqrt{S}}{2}\left(\frac{M_{\rm PBH}|_{T_\phi}-M_{\rm PBH}(t)}{M_{\rm PBH}|_{T_\phi}}\right)^{p-1},
    \end{equation}
where $p$ is a free parameter. $\Gamma_{\rm sc}$ is the semi-classical particle loss rate. The loss rate of PBH mass is then modified by
    \begin{equation}
    \frac{dM_{\rm PBH}}{dt}=S^{-\Delta N}\left(\frac{dM_{\rm PBH}}{dt}\right)_{\rm sc}.
    \end{equation}
We compute the corresponding spectra for {\bf BP}s' in Table \ref{table1} with different $p$ values, as shown in Fig. \ref{continuous}. 
\begin{figure}
    \centering
    \includegraphics[width=0.5\linewidth]{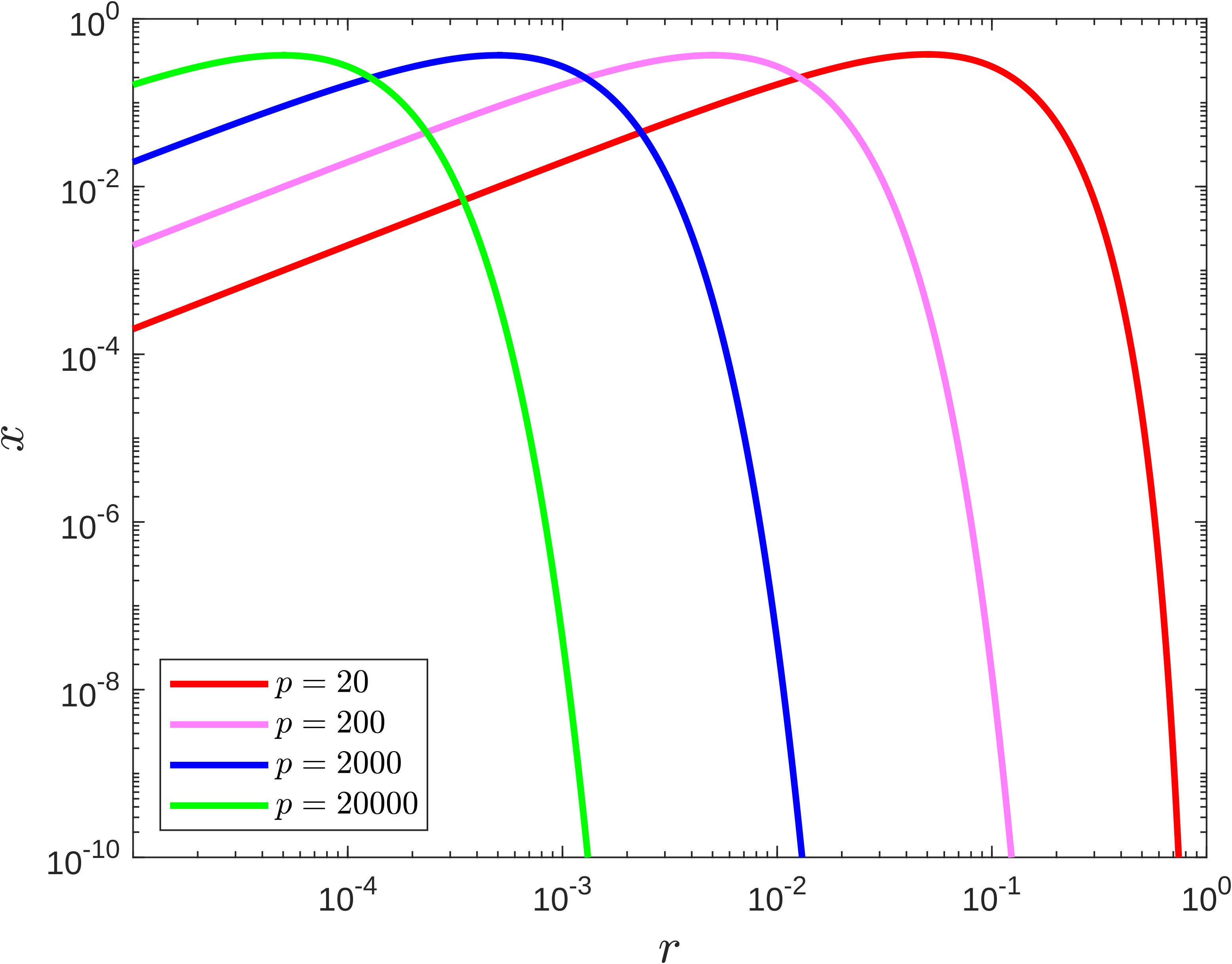}
    \caption{$x-r$ plot}
    \label{x_r}
\end{figure}
Comparing to Fig. \ref{fpbh}, the continuous burden phase transition also exhibits double peaks spectrum, and the $f_{\rm PBH}|_{T_\phi}$ from the constraint of sudden burden phase transition are insufficient for the {\bf BP}s' to reach the experimental limits, except for {\bf BP-4}. Define $r=M_{\rm PBH}(t)/M_{\rm PBH}|_{T_\phi}$, $x=\Delta N/(M_{\rm PBH}|_{T_\phi}\sqrt{\pi G_N})\simeq\Delta N\times(1.23\times10^{-5}\,{\rm g}/M_{\rm PBH}|_{T_\phi})$, then for a fixed $M_{\rm PBH}|_{T_\phi}$, we plot the $x$ value at different ratio $r$ in Fig. \ref{x_r}. As time increases, $r$ reduces from 1 to 0, so the larger the $p$ value is, the slower the PBH enters the burden phase. The $x$ value first raises as $r$ decreases, then gradually falls when $r$ becomes smaller. Therefore, the spectrum of each {\bf BP}s' exhibits a double peaks shape. Also, when $p$ is large, the first peak becomes wider and the second peak amplitude becomes lower because of the delayed entrance of burden phase.

\newpage

\nocite{*}
\bibliographystyle{kp}
\bibliography{bibliography}
\end{document}